# BEYOND THE WELL-MIXED SMOLUCHOWSKI KINETICS: PATTERNS SET CLUSTER SIZE IN MIXTURES OF ATTRACTING PARTICLES

MANUEL DEDOLA[†*], LUDOVICO CADEMARTIRI[†], LARA RIGHI[†]

[†] *Department of Chemistry, Life Sciences and Environmental Sustainability, University of Parma, Parco Area delle Scienze 17A, 43124, Parma, Italy.*

[*] *Author to whom correspondence should be addressed: manuel.dedola@unipr.it*

## ABSTRACT

The Smoluchowski equations for coagulation and fragmentation are usually solved for a well-mixed system, in which every cluster sees the mean concentration. Particles that attract each other break this assumption: they gather into dense regions and leave the rest of the space almost empty. Here we ask how much this changes the cluster population at the same composition and rates. Extending the Keller-Segel model of chemotactic aggregation, we follow the concentration of clusters of each size at every point in space, instead of a single density, and let clusters diffuse, drift toward denser regions up to a packing limit, coagulate and fragment. Depending on the volume fraction, the mixture forms isolated droplets, a bicontinuous labyrinth or inverted bubbles, morphologies also seen in nanocrystal films, chemically attracting colloids and cell networks. Fragmentation changes the cluster size but not the pattern. The pattern, instead, changes the cluster population: by packing mass into dense regions, it raises the mean cluster size in dilute mixtures by 38% at a volume fraction $\phi$ of 0.30 over a well-mixed system, and makes it almost independent of composition.

## INTRODUCTION

Cluster populations are usually described by the Smoluchowski equations,[1] written for a well-mixed system in which every cluster sees the mean concentration. Diffusion smooths out concentration differences and keeps this picture valid. Attracting particles do the opposite: they build dense and empty regions, so coagulation, which depends on the product of local concentrations, runs at different rates in different places. The question of this work is simple: *how are the clusters distributed between the dense and dilute parts of such a mixture, and how much does the whole population differ from the well-mixed prediction at the same composition and rates?* For biomolecular condensates such as P granules, which form by phase separation, a thermodynamic theory shows two things at equilibrium. The dense and the dilute phase have different assembly size distributions, that is, different fractions of molecules in monomers, dimers and larger clusters. Taken together, the two phases can also hold larger assemblies than a uniform mixture with the same composition.[2] In that theory the two phases come from a free energy and are treated as uniform compartments. In our model the pattern comes from transport instead: an attraction between the clusters, limited by volume filling, makes the uniform mixture unstable, as in spinodal decomposition and not by nucleation, and coagulation and fragmentation, not tied by detailed balance, act everywhere in the pattern. This case has not, to our knowledge, been quantified.

Whether a pattern changes the reactions that run inside it also matters in cells, where phase-separated compartments are thought to concentrate reactants and so regulate biochemistry.[3–5] Chemical gradients also drive clustering in experiments. Silver chloride micromotors gather through the ion gradients they release.[6] Catalytic Janus colloids form a dynamic cluster phase that a chemotactic aggregation model reproduces,[7] and in the same kind of system the cluster-size distribution follows from measured aggregation and fragmentation rates,[8] the two processes in our kinetics. Bacteria that secrete their own attractant form comparable spots and stripes,[9,10] and density-dependent motility gives similar stripes in engineered colonies.[11,12]

The attraction in our model has the form of the Keller-Segel model of chemotactic aggregation, in which particles diffuse and drift up the gradient of a chemical signal $v$: $\partial_t u = D\Delta u - \chi\nabla \cdot (u\nabla v)$.[13–15] We replace the signal with the local mass, so particles drift toward the places where there is more mass around them. In the Keller-Segel model, self-attraction can make the density diverge in finite time, because it ignores the finite size of the particles. Volume-filling models remove this divergence by capping the local density, and crowding then produces extended spatial patterns.[15]

Clustering of this kind is also common in soft matter: self-propelled colloids assemble into dynamic "living" crystals,[16] and purely repulsive self-propelled particles separate into dense and dilute phases above a threshold in density and activity.[17,18] In these active fluids, phase separation can stop before the dense phase merges into one large region, leaving dense clusters in a dilute phase, or dilute bubbles in a dense phase, that keep forming and breaking up.[19] Passive systems show similar morphologies: colloidal suspensions whose dense phase relaxes slowly form networks,[20,21] while colloidal particles that arrest the phase separation of two liquids give droplets or bicontinuous gels, the *bijels* (bicontinuous interfacially jammed emulsion gels).[22,23]

In two recent papers, Weyer et al. showed that in Keller-Segel systems that conserve mass the aggregates keep coarsening, unless cell division and death, added as a source term in the density equation, stop them. Both studies point to systems with several species as a next step.[24,25] Their models have a single field, the cell density $\rho(\boldsymbol{r}, t)$, and an aggregate is simply a region where $\rho$ is high. Our model follows one field for each cluster size, $n_k(\boldsymbol{r}, t)$, so it tells how many clusters of each size sit at each point. Each size is a species with its own diffusion and drift, and coagulation and fragmentation turn one species into another.

In this work we study such a mixture with a two-dimensional continuum model in a Eulerian framework, i.e., we do not follow single particles, but the concentration of clusters of each size at each point in space. Clusters diffuse, merge and break, and they attract each other through a non-local attraction that stops where the space is full. Without attraction the mixture stays uniform, and we use it as the reference. With attraction it forms droplets, a labyrinth or bubbles, depending on the volume fraction, and at $\phi = 0.30$ the mean cluster size is 38% larger than in the uniform mixture. Fragmentation changes the cluster size but not the pattern. We keep the reaction rates the same everywhere, so we do not study how the interactions change them. Finally, we compare the morphologies with experiments on nanocrystals, active colloids and cells.

## MATHEMATICAL MODEL

Our model comes from the Keller-Segel model of chemotaxis.[13,14] In that model, particles produce a chemical signal and move toward the places where the signal is stronger. We make three changes. First, there is no signal: particles move toward the places where there is more mass around them, as if they attract each other. Second, we do not follow single particles but clusters of each size, and each size has its own equation; Smoluchowski coagulation and fragmentation link these equations together.[1] Third, each place can hold only a limited amount of mass (volume filling).[15] We keep the model as simple as possible, so that a patterned mixture and a uniform one differ only in where the mass is. The model is two-dimensional and the particles we have in mind are three-dimensional, but they move on a surface and form flat clusters one particle thick, so both their motion and their clusters are two-dimensional. The equations can still be written in three dimensions without changing their form. We divide space into square cells, and in each cell, we follow the concentrations $n_k(\mathbf{r},t)$ of clusters that contain k monomers, with $k = 1, \dots, K_{max}$. A cluster is a bound aggregate that fits inside one cell, and each cell has its own cluster size distribution. The concentrations change in time according to coupled advection-diffusion-reaction equations:

$$\partial_t n_k = D_k \Delta n_k - \nabla \cdot (n_k \mathbf{v}_k) + \mathcal{R}_k \qquad [1]$$

The first term is diffusion: larger clusters diffuse more slowly, following the Stokes-Einstein scaling $D_k = D_1 k^{-1/d_f}$ for aggregates of fractal dimension $d_f$,[26] which gives $D_k = D_1/\sqrt{k}$ for compact two-dimensional clusters ($d_f = 2$). The second term describes the drift caused by the attraction; the drift velocity $\mathbf{v}_k$ is driven by a non-local interaction field $\tilde{M}(\mathbf{r},t)$, a short-range average of the aggregate mass $M(\mathbf{r},t) = \sum k\, n_k$. Mathematically, $\tilde{M}$ is calculated by using a spatial Gaussian convolution over the domain $\Omega$, a square of side L with periodic boundaries:

$$\tilde{M}(\mathbf{r},t) = \int_\Omega \frac{1}{2\pi\sigma^2} \exp\left(-\frac{|\mathbf{r}-\mathbf{r}'|^2}{2\sigma^2}\right) M(\mathbf{r}',t)\, d\mathbf{r}' \qquad [2]$$

where $\sigma = 2$ sets the range of the interaction. Because the Gaussian decays fast, only cells within 4σ of $\mathbf{r}$ contribute. The drift velocity is:

$$\mathbf{v}_k = \chi_k \left[1 - \left(\frac{M}{M_{max}}\right)^2\right] \nabla \tilde{M}(\mathbf{r}, t) \qquad [3]$$

where $\chi_k = \chi/\sqrt{k}$ is the size-dependent drift coefficient. Diffusion and drift are both limited by the friction of the moving cluster, which grows as $k^{1/2}$ for a compact two-dimensional cluster and as $k^{1/3}$ for a compact three-dimensional one. Taking the attractive force on a cluster as independent of its size, both coefficients therefore decrease by the same factor, and their ratio, $\frac{\chi_k}{D_k} = \frac{\chi}{D_1}$, is the same for every cluster size. This ratio is the analogue of the chemotactic Péclet number,[27] $Pe = \chi/D_1$. It measures the attraction against diffusion and is the main control parameter of this work. $Pe$ has units of inverse density. With densities counted in monomers per cell it is a pure number, and the group that does not depend on this choice is $Pe \cdot M_{max}$, equal to 144 at $Pe = 12$. If the force grew with cluster size, as it can between attracting particles, larger clusters would have a larger Péclet number. With $\chi_k$ independent of k or growing as $\sqrt{k}$, the pattern wavelength and the increase in mean cluster size change by about 1% or less (Table S3). $M_{max}$ is the largest number of monomers a cell can hold, so $M/M_{max}$ is the fraction of the cell that is occupied. The crowding factor in Eq. 3 is close to 1 in nearly empty cells and becomes zero when a cell is full, so no more mass can enter it by drift. This limits the density, which in the classical Keller-Segel model can diverge in finite time.[13,15] The volume fraction is $\phi = \bar{M}/M_{max}$, where $\bar{M}$ is the mean number of monomers per cell over the whole domain. Cells that hold more monomers than this mean, $M > \bar{M} + \epsilon$ with a small margin $\epsilon$, form the dense phase, and the other cells form the dilute phase. For a fixed value of $\phi$, varying $M_{max}$ is equivalent to rescaling $Pe$ and the ratio $K_0/\gamma$, both of which we vary directly.

The drift in Eq. 3 describes an attraction between the particles. A pair potential, such as Lennard-Jones between molecules or the van der Waals attraction between nanoparticles described by Hamaker,[28] acts between single particles, which we do not follow. When one follows densities, an attraction of range $\sigma$ becomes a drift toward the regions where the locally averaged density, here $\widetilde{M}$, is higher.

The last term $\mathcal{R}_k$ describes aggregation and fragmentation with mass-action kinetics:

$$\mathcal{R}_k = \frac{1}{2} K_0 \sum_{i=1}^{k-1} n_i \, n_{k-i} - K_0 n_k \sum_{i=1}^{K_{max}-k} n_i - \gamma n_k (1 - \delta_{k,1}) + \delta_{k,1} \gamma \sum_{j=2}^{K_{max}} j \, n_j \qquad [4]$$

Its first two terms are the discrete Smoluchowski coagulation equation[1] with a single rate constant $K_0$, so that all clusters are equally reactive; to these we add complete fragmentation of each cluster into its monomers. From left to right: a cluster of size k forms when two smaller clusters merge; it disappears when it merges with another cluster, with the sum truncated at $K_{max} - k$ so that no cluster exceeds the largest size; it breaks up at rate γ, a loss that excludes monomers through the factor $1 - \delta_{k,1}$; and each cluster of size j that breaks releases j monomers, which appear only in the equation for $k = 1$ through the Kronecker delta $\delta_{k,1}$. We take $K_{max} = 10$, below the cell capacity $M_{max} = 12$, since a cluster must fit inside one cell. These are the simplest choices. A surface energy would make fragmentation depend on cluster size and link it to coagulation through detailed balance; we leave it out. Smoluchowski kinetics can be unstable on its own. If the coagulation rate grows faster than linearly with cluster size, a gel forms in finite time.[29] Gelation cannot occur in our model, because the rate constant $K_0$ is independent of cluster size and cluster size is limited to $K_{max}$. The patterns therefore come from attraction arrested by volume filling in physical space. We keep $K_0$ and $\gamma$ the same in every cell, so any change in cluster size comes only from where the mass is. In a real system, the interactions in full cells would also change the rates through the chemical potentials;[2] we do not include this effect.

## SIMULATION PARAMETERS AND NUMERICAL IMPLEMENTATION

**Numerical method.** We solve the equations with a finite-volume method. At each cell face, the drift flux is computed from the cell the clusters come from, using a linear profile inside that cell limited by the *minmod* function.[30,31] Time is advanced with the explicit Euler method. When a cell is almost full, we reduce the flux of clusters that enter it, so the total mass is conserved and the local mass exceeds $M_{max}$ by at most 2% (Table S5). Where M exceeds $M_{max}$, the crowding factor in Eq. 3 is set to zero. A time step half as large gives the same results. On a finer grid ($\Delta x = 0.5$, with the interaction range still $\sigma = 2$), the pattern is almost the same and the increase in mean cluster size is slightly larger, for example 42% instead of 38% at $\phi = 0.30$ and $\gamma = 0.5$. The values reported here are therefore conservative (Table S5).

**Parameters.** Lengths are in units of the grid spacing ($\Delta x = 1$). Time is in simulation units, in which $D_1 = 0.3$. Simulations start from a uniform monomer density, $n_1 = \phi\, M_{max}$, with a random perturbation of $\pm 0.25$. Unless stated otherwise (Table S1), simulations use $K_{max} = 10, M_{max} = 12, \sigma = 2, K_0 = 0.8$ and $d\tau = 10^{-3}$ on periodic $L \times L$ domains with $L = 50$. The interaction range $\sigma$ sets the size of the pattern (Fig. S9, Table S6). At $\phi = 0.50$ and $\gamma = 0.5$, when $\sigma$ goes from 1.5 to 3, $\lambda$ goes from 7.9 to 13.4. The dense phase still fills up to the capacity and covers the same area (0.47-0.48). The mean cluster size grows by 20% at $\sigma = 1.5$ and $2$, and by 17% at $\sigma = 3$, because in this case the rest of the domain does not become completely empty.

**Observables.** The maps and averages in the results are built from two moments of the local size distribution, the local mass and the local number of clusters, and from their ratio:

$$M(\mathbf{r},t) = \sum_{k=1}^{K_{\max}} k\, n_k(\mathbf{r},t)\,, \qquad N(\mathbf{r},t) = \sum_{k=1}^{K_{\max}} n_k(\mathbf{r},t)\,, \qquad \langle k\rangle(\mathbf{r},t) = \frac{M(\mathbf{r},t)}{N(\mathbf{r},t)}$$

$M(\mathbf{r},t)$ counts the monomers in a cell, free or bound in clusters, whereas $N$ counts the objects in it, free monomers and clusters of any size alike; with $\Delta x = 1$ both are numbers per cell. Their ratio $\langle k\rangle$ is the number-average cluster size. Cells with $M < 0.1$ are shown in grey in the $\langle k\rangle$ maps. In the maps, the droplets, the labyrinth and the bubbles are made of many cells, not single clusters. The spatial structure is quantified by the radially averaged structure factor $S(|\boldsymbol{q}|)$ of the mass fluctuations $\delta M(\mathbf{r})$ (Eqs. S1-S4).

**Crowding.** The limit $M_{max}$ applies to the total number of monomers in a cell, not to single clusters: a cluster of 4 and one of 3 fit together, two clusters of 8 do not. We thus assume that clusters pack without gaps and never jam. Only the drift is limited, because it is the drift that pushes clusters into the dense phase, while diffusion moves them out. Painter and Hillen use the linear form $\left[1 - \frac{M}{M_{max}}\right]$,[15] which describes rigid blocks, while in volume-filling models larger exponents describe objects that can still squeeze into the free space between them.[32] We use the quadratic form so that crowding barely slows the clusters in cells far from full and stops them only when a cell is close to full. In nanoparticle dispersions, crowding also slows down the collisions between large clusters, even at low volume fractions.[33] Here we include only its effect on transport.

**Phase classification.** The radial structure factor $S(|\boldsymbol{q}|)$ (Eqs. S1 to S4) is used to determine the characteristic length scale of the patterns on the periodic domain. Topological information can be obtained by considering percolation. To account for the structures that wrap around the torus we use a 2x2 tiling and label connected regions on the enlarged grid. This avoids the problem of identifying if a region connects both sides of the domain. A phase percolates when the bounding box of a connected region spans at least L in width or height. A fixed density threshold would tie the classification to the composition we are varying, so we use the dense and dilute phases defined in the Mathematical Model, with $\epsilon = 0.5$ set above the initial noise amplitude.

This gives four states. In a uniform mixture, nothing crosses the threshold. Isolated droplets form when the dense phase exists but does not percolate. Inverted bubbles form when the dense phase percolates and the dilute phase is fragmented. In a bicontinuous labyrinth, both phases percolate.

## RESULTS AND DISCUSSION

**Pattern formation.** Figure 1 shows how the patterns form. We start from a uniform mixture of monomers with small random fluctuations. The fluctuations grow, and the structure factor develops a peak whose height rises by more than four orders of magnitude in about five time units (Fig. 1C, D). By $\tau \approx 5$ the dense phase reaches the capacity $M_{max}$, and the peak settles at $\lambda \approx 10$ lattice units, the same for all compositions. The rest of the domain is nearly empty, so the dense phase covers a fraction of the domain close to $\phi$ (0.48 at $\phi = 0.50$; Table S2).

After $\tau \approx 5$ the growth slows sharply, and between $\tau = 10$ and $30$ the structure factor barely changes (Fig. 1C). The pattern still coarsens slowly, and the maps at $\tau = 200$ show larger domains than at $\tau = 10$ (Fig. 1B, Fig. S5). Slow coarsening that does not stop is expected for chemotactic aggregates that conserve mass and saturate into plateaus.[24,25] Our reactions conserve mass, so they cannot stop it.

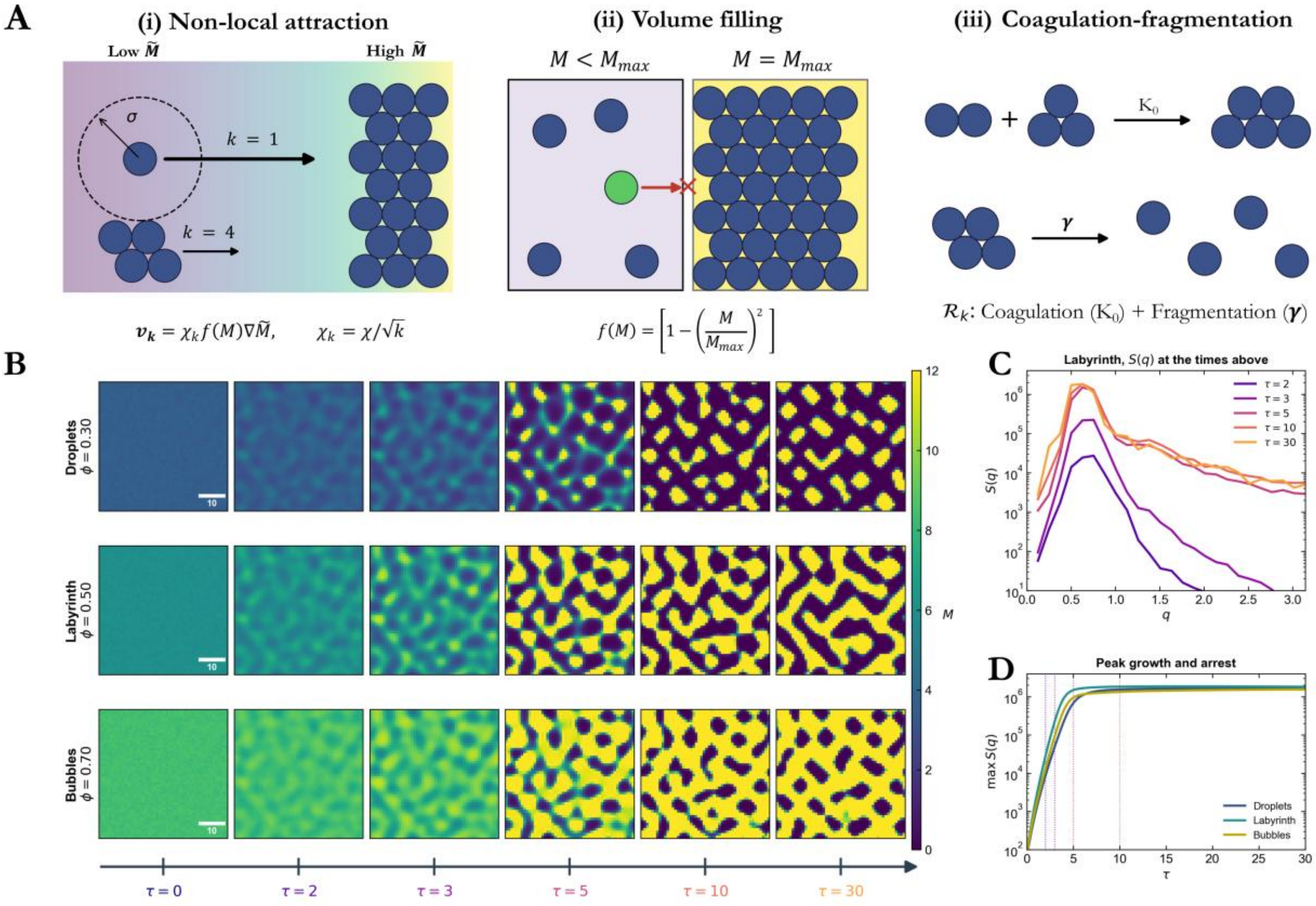

**Figure 1. Dynamics of model pattern formation.** (**A**) The model combines size-dependent attraction, volume filling and coagulation-fragmentation. (**B**) Total mass $M(\boldsymbol{r})$ in each cell, from a uniform state into droplets ($\phi = 0.30$), labyrinth ($\phi = 0.50$) and bubbles ($\phi = 0.70$). The maps show concentration fields, not single particles. (**C, D**) Radial structure factor $S(|\boldsymbol{q}|)$ of the labyrinth (C) and its peak height (D): fast growth until $\tau \approx 5$, then slow coarsening. $Pe = 12, \gamma = 0.05, K_{max} = 10, M_{max} = 12, \sigma = 2, L = 50$ (scale bars: 10 lattice units).

**Morphology.** Figure 2 maps the morphology at $\tau = 200$ over the $(\phi, Pe)$ plane, Fig. S5 shows every state, and Figs. S2-S4 show the full analysis of one state per regime at $\tau = 10$. Without attraction ($Pe = 0$) the mixture stays uniform (Fig. S1 and bottom row of Fig. S5). For $Pe \geq 2$ every state forms a saturated pattern by $\tau = 200$, except the most dilute one ($\phi = 0.10$ at $Pe = 2$), which is still growing. In the range studied, $Pe$ therefore mainly sets how fast the pattern forms, not whether it forms.

The volume fraction sets the morphology. For $Pe \geq 10.5$ we find isolated droplets for $\phi \leq 0.35$, a bicontinuous labyrinth for $0.42 \leq \phi \leq 0.54$ and inverted bubbles for $\phi \geq 0.60$. At lower $Pe$ both boundaries move: the droplets extend to $\phi = 0.42$ for $Pe \leq 8$, and at $Pe = 3$ and 4 the state at $\phi = 0.54$ is classified as bubbles. Close to $\phi = 0.50$ the two phases have similar areas, so whether the dilute phase percolates can depend on small details of the pattern. The reason is geometric: dense cells fill up to $M_{max}$ and the rest of the domain is nearly empty, so $\phi$ fixes the area of the dense phase, and the area decides which phase percolates. Both phases percolate only when their areas are comparable, which is also where the structure factor is highest.

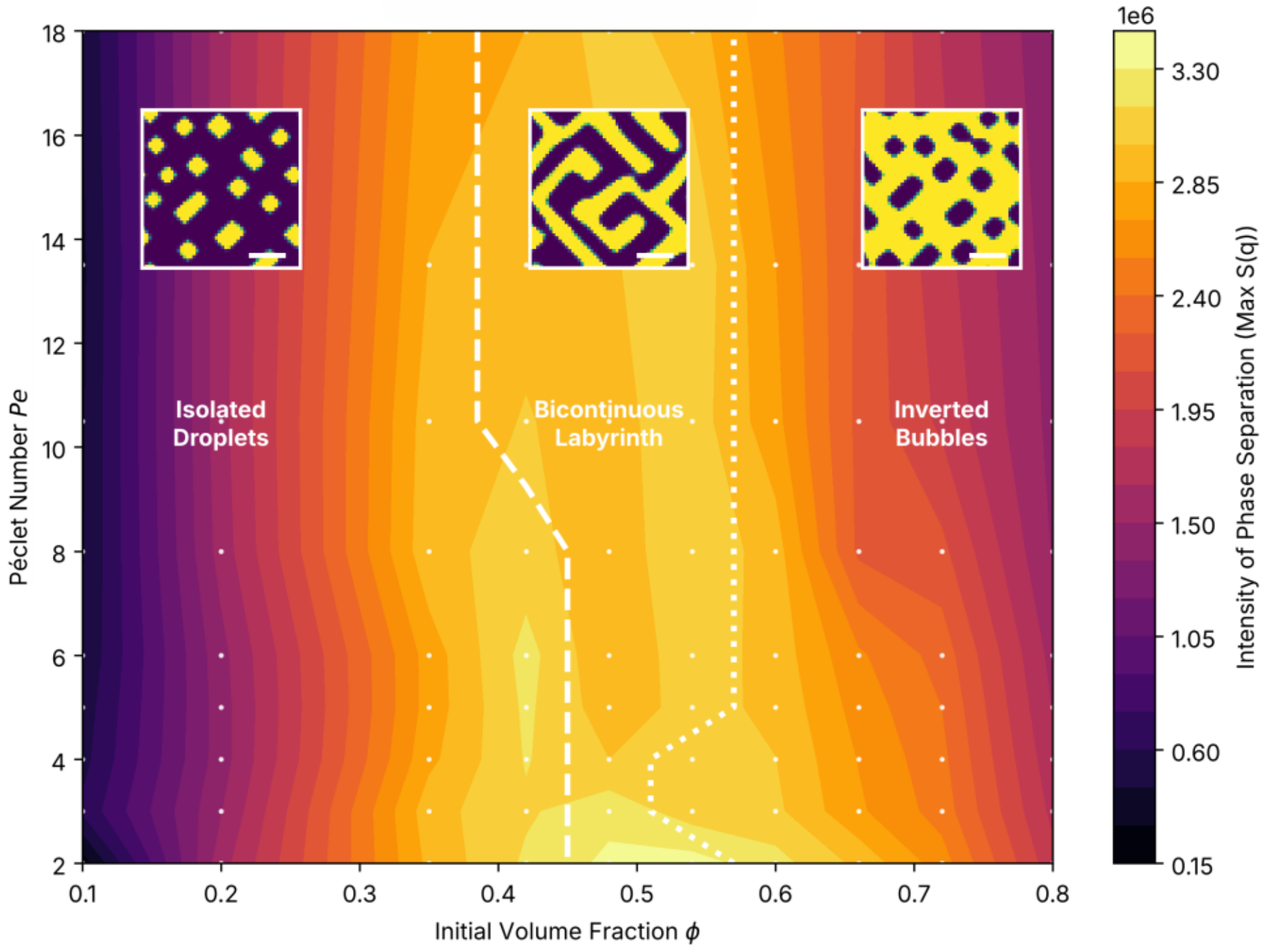

**Figure 2. Morphology phase diagram in the $(\phi, Pe)$ plane at $\tau = 200$, $\gamma = 0.05$, $K_{\max} = 10$, $M_{\max} = 12$, $\sigma = 2$.** The color map shows the structure factor peak, $\max S(q)$. Insets show representative $M(\boldsymbol{r})$ fields for each labeled regime (scale bars: 10 lattice units).

**Clusters inside the pattern.** Figure 3 looks inside the labyrinth ($Pe = 12, \phi = 0.50$) at $\gamma = 0.5, 2$ and $5$. The mass maps look almost the same for the three values of $\gamma$: both phases still percolate, $\lambda$ stays between 9.4 and 10.2 lattice units and the occupied area between 0.475 and 0.486 (Fig. 3A-C, Table S2). The clusters, instead, follow γ: faster fragmentation lowers the mean cluster size over the domain from 3.10 to 2.01 and 1.55 (Fig. 3E-G, Table S2).

The pattern itself changes the cluster population. Without attraction ($Pe = 0$) the same mixture stays uniform and reproduces the well-mixed Smoluchowski limit, since every cell holds the same mass. At the same $\phi, \gamma$ and time its mean cluster size is 2.59, 1.69 and 1.35. The patterned mixture therefore holds clusters 15-20% larger on average (Table S4). The difference appears when the pattern forms, between $\tau \approx 2.5$ and $4$, and then stays nearly constant (Fig. 3H); by $\tau = 200$, after the pattern has coarsened, it is slightly larger (41%, 22% and 10% at $\phi = 0.30, 0.50$ and $0.70$, against 38%, 20% and 9% at $\tau = 10$, for $\gamma = 0.5$; Fig. S8). Coagulation depends on the product of local concentrations, so packing the same mass into the occupied branches speeds it up there and shifts the whole population toward larger clusters, while fragmentation acts at the same rate everywhere.

The size of the effect depends on composition (Fig. 3D, Table S4). It is largest in dilute mixtures, where the pattern raises the local concentration the most: at $\gamma = 0.5$ the increase is 38% at $\phi = 0.30$, 20% at $\phi = 0.50$ and 9% at $\phi = 0.70$, and it grows slightly on finer grids (Table S5). In the patterned mixture, instead, the mean cluster size hardly depends on $\phi$: from $\phi = 0.30$ to $0.70$ it changes by at most 4% at every $\gamma$, against 18-31% in the uniform mixture. Occupied cells always fill to capacity, so adding mass makes the occupied area larger and changes the morphology but leaves the cluster population inside it the same. Indeed, at $Pe = 12$ the well-mixed Smoluchowski equations solved at $M = M_{max}$ give a mean cluster size within 6% of the patterned one at every $\gamma$ (Table S4). This also gives an upper limit. If all the mass sat in full cells, the increase at γ = 0.5 would reach about 46% at $\phi = 0.30$, 24% at $\phi = 0.50$ and 12% at $\phi = 0.70$. As the pattern coarsens, the partly filled boundary between the two phases gets smaller, and the increase slowly moves toward this limit (Fig. S8). The reactions follow the local concentration rather than the mean one, as proposed for biomolecular condensates,[4,5] but here the pattern builds its own compartments.

At $\gamma = 0.05$, used for the morphology runs, fragmentation is slower than pattern formation ($1/\gamma = 20$) and the largest clusters dominate ($\langle k \rangle = 5.85$ over the domain). The pattern raises the mean cluster size by 14% here as well (Table S4), although at this rate the populations are still evolving at $\tau = 10$. At low γ many clusters reach the largest size $K_{max}$. At $\gamma = 5$ only a small fraction of the mass does, and the pattern still raises the mean cluster size by 8-24% (Table S4), so the effect is not caused by the size limit. The results do not depend on the noise realization or on the box size (Table S2, Fig. S7). They also do not depend on how the run starts: starting from the clusters of the uniform mixture instead of monomers gives the same values (Fig. S10). The constant and Brownian kernels give the same mass map and mean cluster sizes within about 1% (Fig. S6).

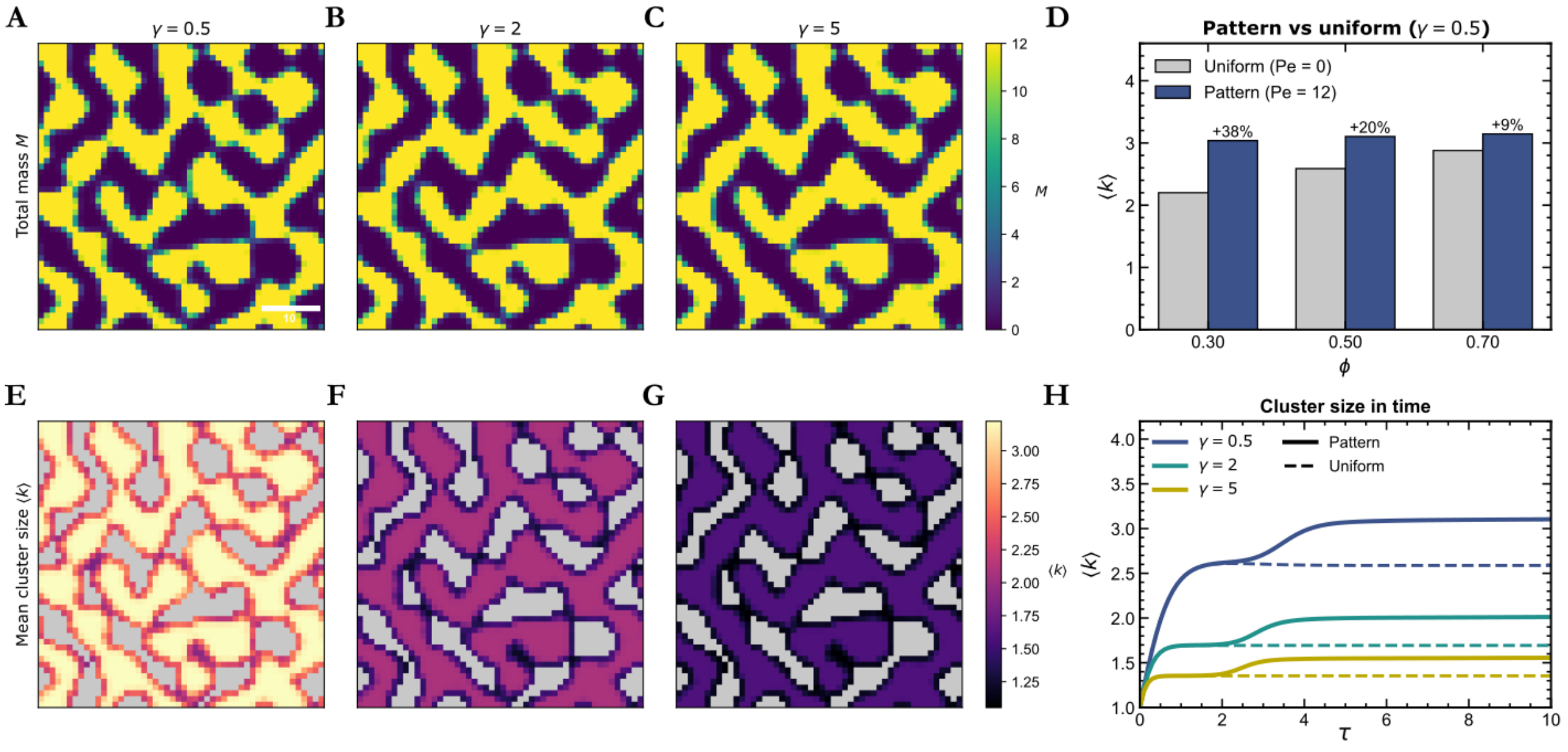

**Figure 3. The pattern sets the cluster population. (A-C)** Total mass $M(r)$ and (**E-G**) mean cluster size $\langle k \rangle = M/N$ of the labyrinth ($Pe = 12, \phi = 0.50, K_{max} = 10, \tau = 10$) at $\gamma = 0.5, 2$ and 5: the pattern does not change with γ, the cluster size does. Grey cells have $M < 0.1$ (< 0.1% of the total mass); scale bar: 10 lattice units. (**D**) Mean cluster size over the whole domain for the uniform mixture ($Pe = 0$) and the patterned mixture ($Pe = 12$) at $\phi =$

$0.30, 0.50$ and $0.70$ ($\gamma = 0.5$); percentages give the increase due to the pattern (all $\gamma$ in Table S4). (**H**) Mean cluster size over the whole domain versus time at $\phi = 0.50$, with the pattern (solid) and without it (dashed).

The structure is set by transport and volume filling ($Pe$ and $\phi$), and the cluster population by the kinetics (γ) together with the structure, which decides how much mass sits in each place. Because $D_k$ and $\chi_k$ scale in the same way, transport does not sort clusters by size, so changing $\gamma$ moves mass between sizes but not between places. Even when transport does sort them ($\chi_k \propto \sqrt{k}$, Table S3), the result hardly changes, because the dense phase is filled to capacity anyway. This is the opposite of chemically active droplets, where the reactions that make and destroy droplet material set the droplet size.[3] Here the reactions only move mass between cluster sizes, and it is the pattern that shapes them.

**Comparison with experiments.** The same morphologies appear in real systems, together with the ingredients of our model. The closest case is PbSe nanocrystals on graphite: islands at 4% coverage, a labyrinth at 41% and a layer with holes at 68% (Fig. 4D-F, compare Fig. 4A-C), which Ou et al. read as phase separation of attracting nanocrystals frozen by solvent evaporation.[34] The same ingredients are found in active systems where the attraction comes from chemical signals. Au-Pt Janus colloids in hydrogen peroxide attract each other through the chemical gradients they produce,[7] and at area fractions of 5-10% they form isolated clusters whose size distribution follows from measured aggregation and fragmentation rates.[8] Photoactivated hematite particles attract silica spheres by diffusiophoresis and glue them into space-filling gels.[35] Endothelial cells guided by chemotaxis form a network whose branches vary in thickness, unlike our uniform dense phase. Its topology, however, changes with density as in our model: separate groups of cells, then a connected network, then a layer with holes. Serini et al. reproduce it with a chemotaxis model whose signal range and pressure term play the roles of our smoothing kernel and volume filling.[36] The PbSe nanocrystals, which are passive, show that activity is not needed: the same morphologies appear whenever mass is conserved and an attraction toward denser regions is limited by packing, as in Eq. 3. Our model predicts that patterned samples hold larger clusters than well-mixed ones of the same composition. The model is phenomenological, so the comparison concerns ingredients and morphologies, not numbers. All these systems are close to two-dimensional, as in our model. In three dimensions percolation is easier, so the morphology boundaries would move.

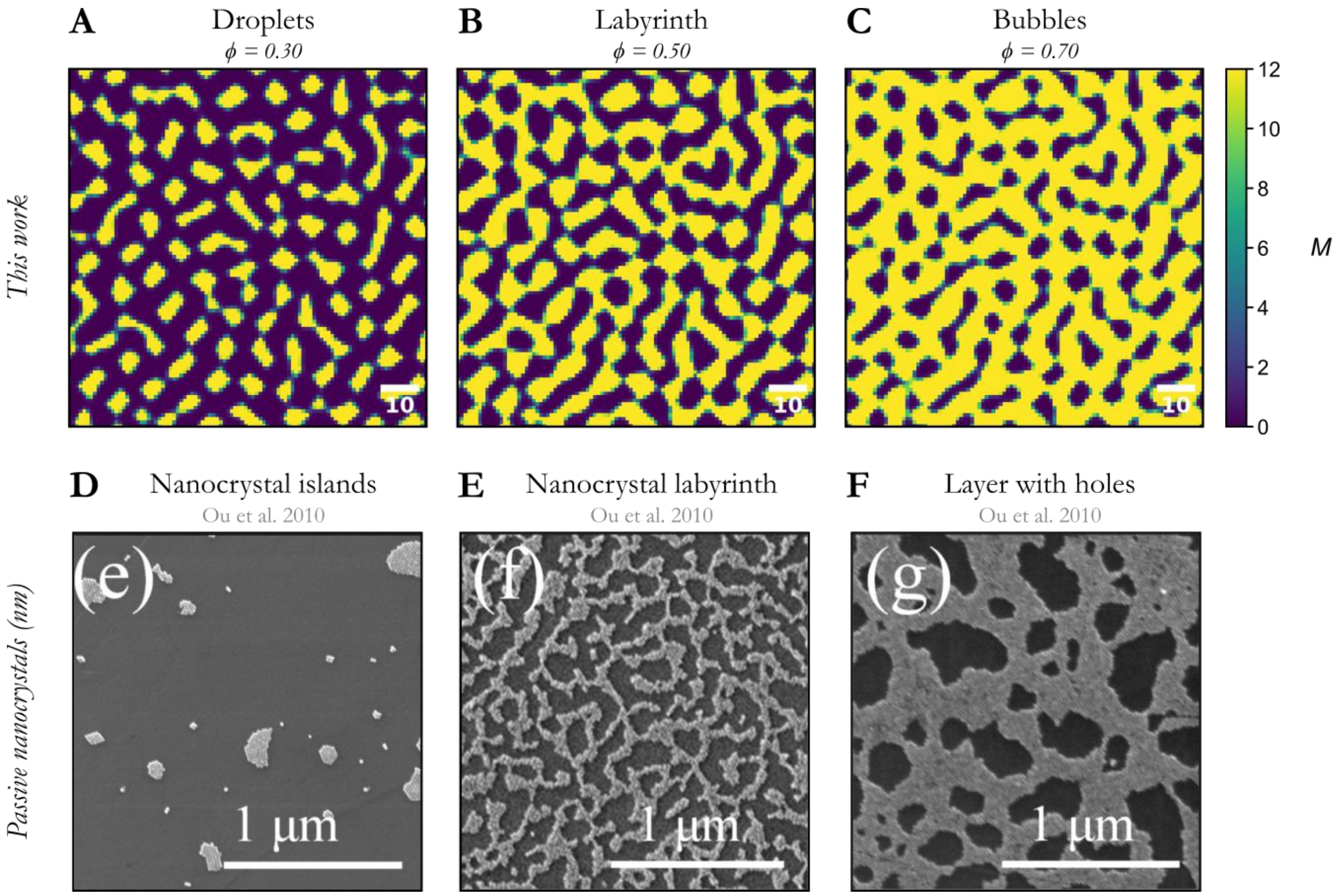

**Figure 4. The same morphologies in the model and in a passive nanocrystal film.** (**A-C**) Model, $M(\boldsymbol{r})$ at $\phi = 0.30, 0.50$ and $0.70$ ($Pe = 12, \gamma = 0.05, \tau = 10, L = 100$; scale bars: 10 lattice units). (**D-F**) PbSe nanocrystals (core 14.6 nm) dried from toluene on graphite at 50 °C, at 4%, 41% and 68% coverage (scale bars: 1 μm).[34] Used with permission of the American Chemical Society, from Interplay of Spinodal and Diffusion-Limited Aggregation in Formation of Nanocrystal Assembled 2D Islands, Ou, Y.-C.; Zhuravlev, K. S.; Fang, J.; Jian, W.-B., J. Phys. Chem. C, 114, 2010 (Fig. 1e-g); permission conveyed through Copyright Clearance Center, Inc.

## CONCLUSIONS

Patterns change the cluster population of the mixture in which they form. Coupling a non-local attraction limited by packing to coagulation and fragmentation, we find three results. First, the pattern packs mass into dense regions and raises the mean cluster size of dilute mixtures by 38% over a well-mixed system with the same composition and rates, and this increase grows slowly toward about 46% as the pattern coarsens. Second, because the dense phase is always full, the cluster size in the patterned mixture hardly depends on composition. Third, fragmentation changes the cluster size but not the pattern. The morphologies (isolated droplets, bicontinuous labyrinth and inverted bubbles) are generic and also appear without attraction, for example with purely repulsive active particles;[19] what is new here is that they change the reactions inside them. The well-mixed Smoluchowski equations underestimate the cluster size in systems that form patterns, but they give the right value if solved at the concentration of the full dense phase. This can be tested in active colloids, where cluster sizes are already measured by microscopy,[8] by comparing patterned and well-mixed samples of the same composition. Our reaction rates are the same everywhere; how the interactions inside full cells change them is the next question. Because the pattern changes the cluster size without changing the composition or the rates, where particles gather becomes a way to control size in materials that grow by joining smaller units, such as nanocrystals that grow by coalescence.[33]

## DATA AVAILABILITY

The equations, the numerical scheme and all parameters needed to reproduce the results are given in the article and in the Supporting Information.

## AUTHOR CONTRIBUTIONS

M.D.: Conceptualization, Methodology, Software, Investigation, Formal analysis, Visualization, Writing (original draft). L.C., L.R.: Writing (original draft), Formal analysis, Visualization, Writing (review and editing).

## SUPPORTING INFORMATION

Model parameters, analysis of single states, robustness to noise, domain size and initial state, kernel comparison, numerical convergence, interaction range, and structure factor definitions.

## NOTES

The authors declare no competing financial interest.

## ACKNOWLEDGEMENTS

This research received no specific grant from any funding agency.

# BEYOND THE WELL-MIXED SMOLUCHOWSKI KINETICS: PATTERNS SET CLUSTER SIZE IN MIXTURES OF ATTRACTING PARTICLES

MANUEL DEDOLA†*, LUDOVICO CADEMARTIRI†, LARA RIGHI†

*† Department of Chemistry, Life Sciences and Environmental Sustainability, University of Parma, Parco Area delle Scienze 17A, 43124, Parma, Italy.*

** Author to whom correspondence should be addressed: manuel.dedola@unipr.it*

## SUPPORTING INFORMATION

**Table S1. Model parameters.** Values used in this work and their role.

| Symbol | Value | Meaning |
|---|---|---|
| $L$ | 50 (100 in Fig. 4; 100, 150 in Fig. S7) | Domain side (periodic) |
| $\Delta x, d\tau$ | 1, $10^{-3}$ | Grid spacing, time step |
| $\tau$ | 10, 200 (Figs. 1B, 2, S5, S8) | Final time |
| $D_1$ | 0.3 | Monomer diffusivity |
| $Pe = \frac{\chi}{D_1}$ | 12 (2-18 in Fig. 2; 0 for the uniform mixture) | Péclet number |
| $\sigma$ | 2 (1.5, 3 in Fig. S9) | Interaction range |
| $M_{max}$ | 12 | Cell capacity |
| $K_{max}$ | 10 | Largest cluster |
| $K_0$ | 0.8 | Coagulation rate |
| $\gamma$ | 0.05 (Figs. 1, 2, 4); 0.5, 2, 5 (Fig. 3) | Fragmentation rate |
| $\phi$ | 0.30, 0.50, 0.70 (0.10-0.80 in Fig. 2; 0.40 in Fig. S1; 0.45 in Fig. S6, 1.00 in Table S4) | Volume fraction |
| – | ±0.25 | Initial noise on $n_1$ |

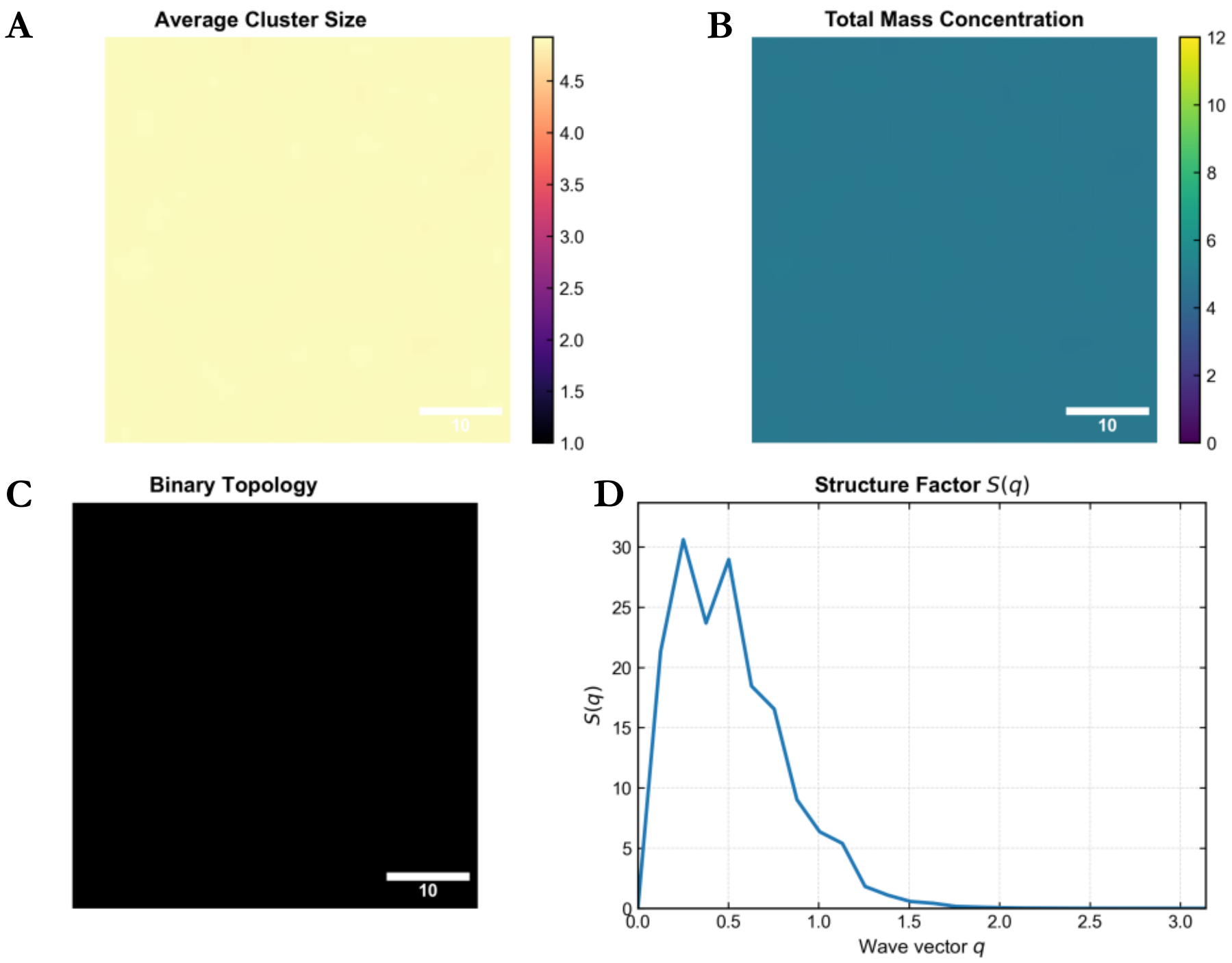

**Figure S1. Uniform mixture without attraction ($Pe = 0, \phi = 0.40, \gamma = 0.05$) at $\tau = 10$.** (**A**) Mean cluster size $\langle k \rangle$. (**B**) Total mass M. (**C**) Dense phase ($M > \bar{M} + 0.5$): no cell is above the threshold. (**D**) Structure factor $S(|\boldsymbol{q}|)$. Scale bars: 10 lattice units.

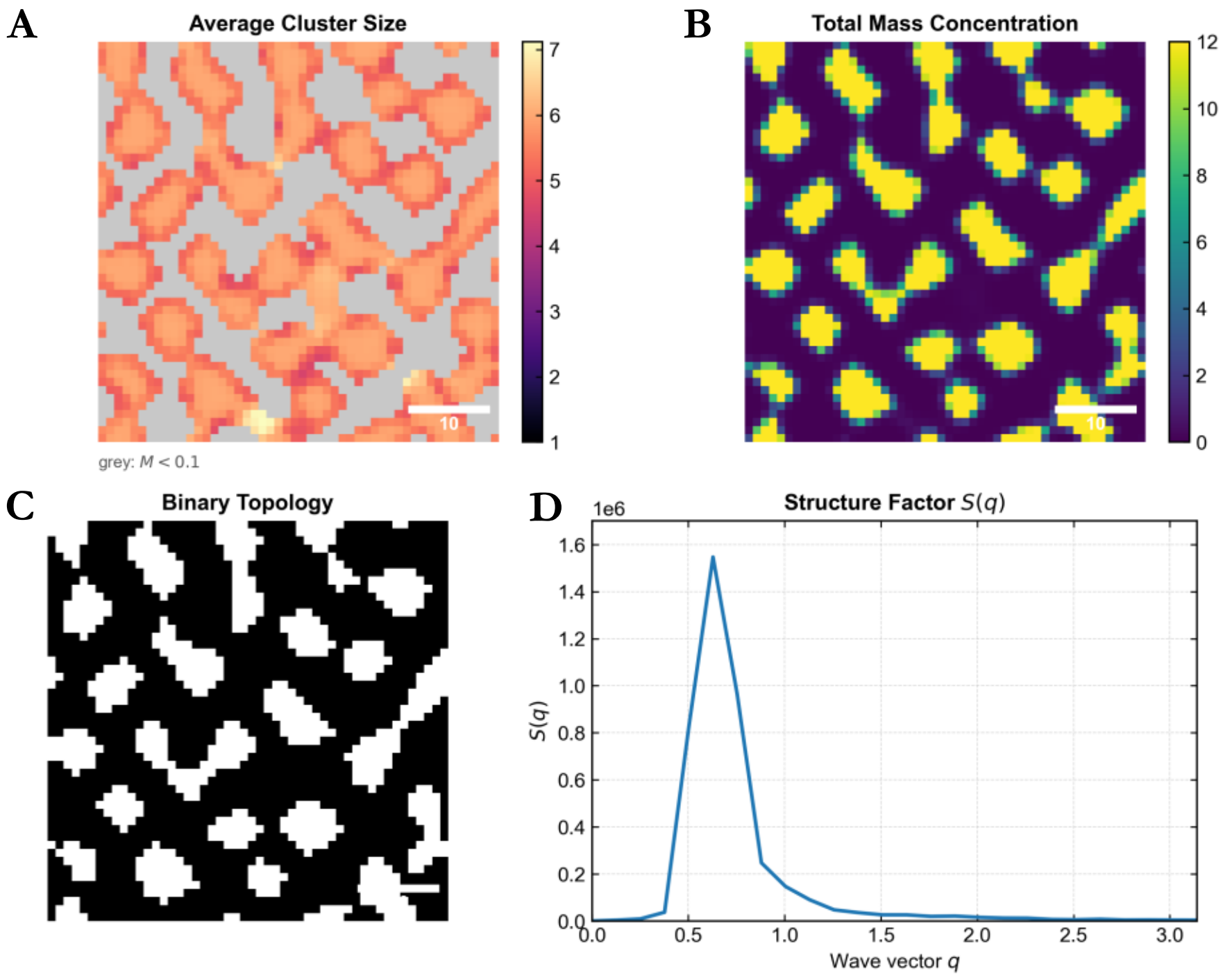

**Figure S2.** ***Droplets* ($Pe = 12, \phi = 0.30, \gamma = 0.05$) at $\tau = 10$.** (**A**) Mean cluster size $\langle k \rangle$; grey cells have $M < 0.1$. (B) Total mass M. (**C**) Dense phase (white, $M > \bar{M} + 0.5$). (D) Structure factor $S(|\boldsymbol{q}|)$. Scale bars: 10 lattice units.

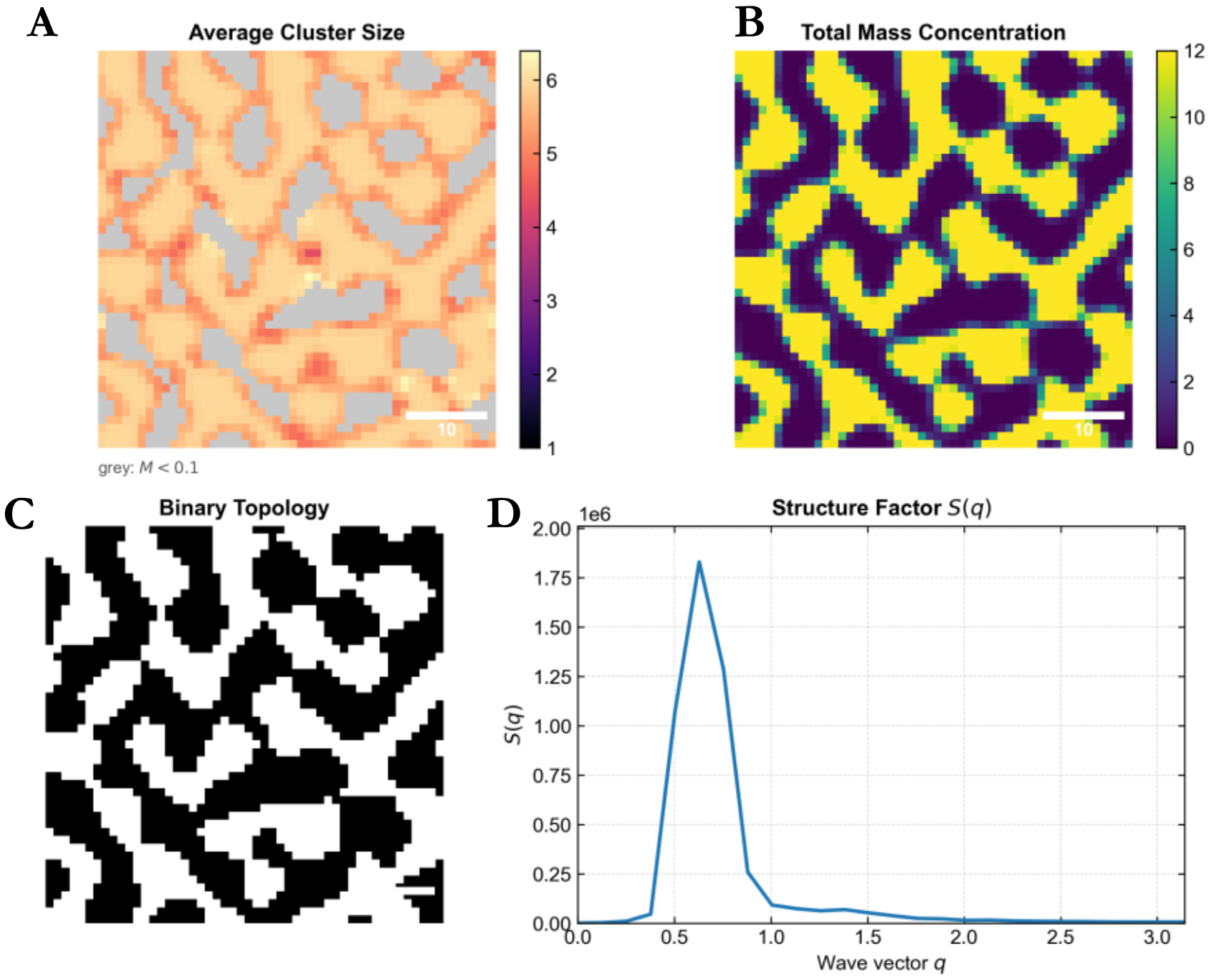

**Figure S3.** ***Bicontinuous labyrinth* ($Pe = 12, \phi = 0.50, \gamma = 0.05$) at $\tau = 10$. Panels as in Fig. S2.**

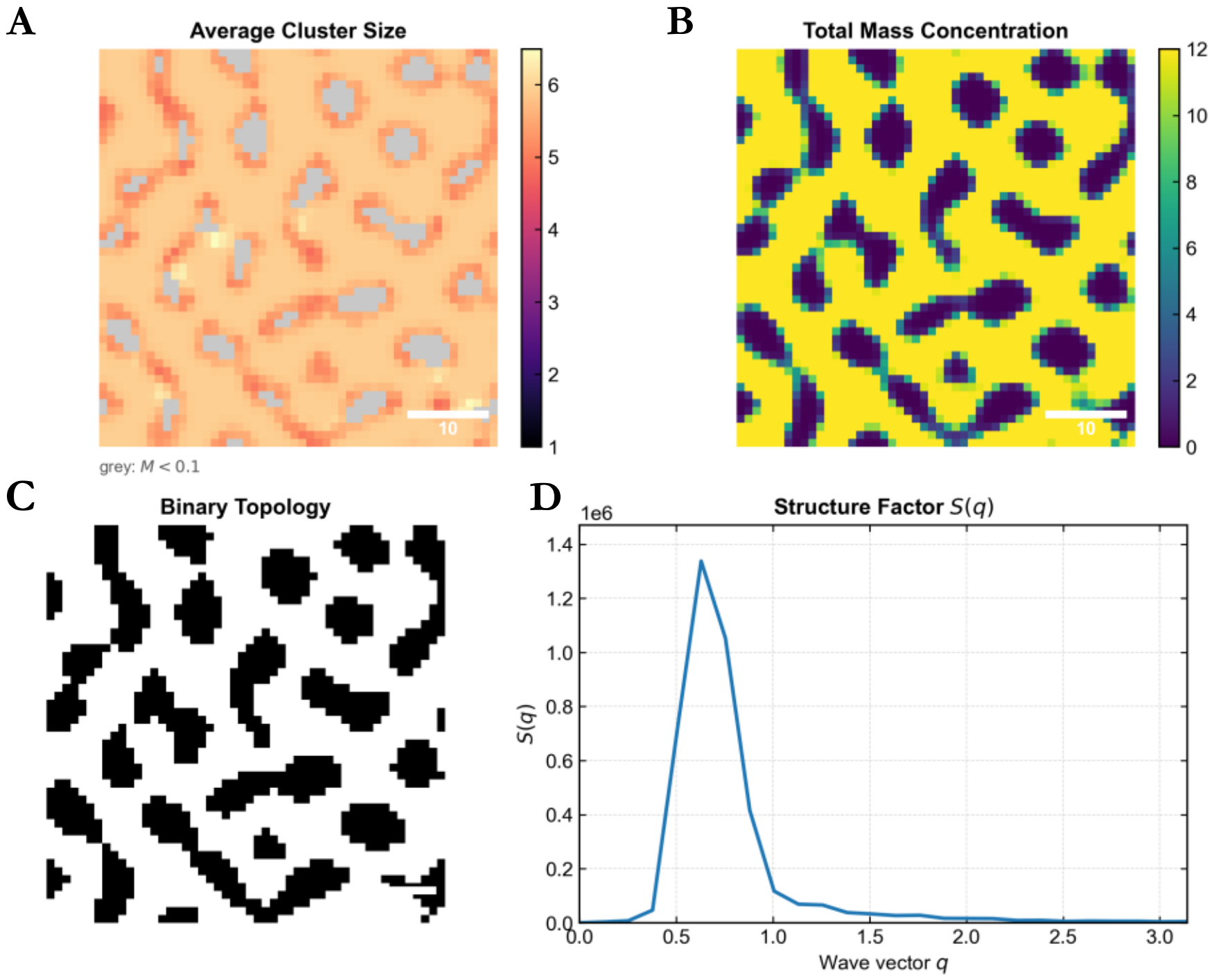

**Figure S4.** *Inverted bubbles* ($Pe = 12, \phi = 0.70, \gamma = 0.05$) *at* $\tau = 10$. *Panels as in Fig. S2.*

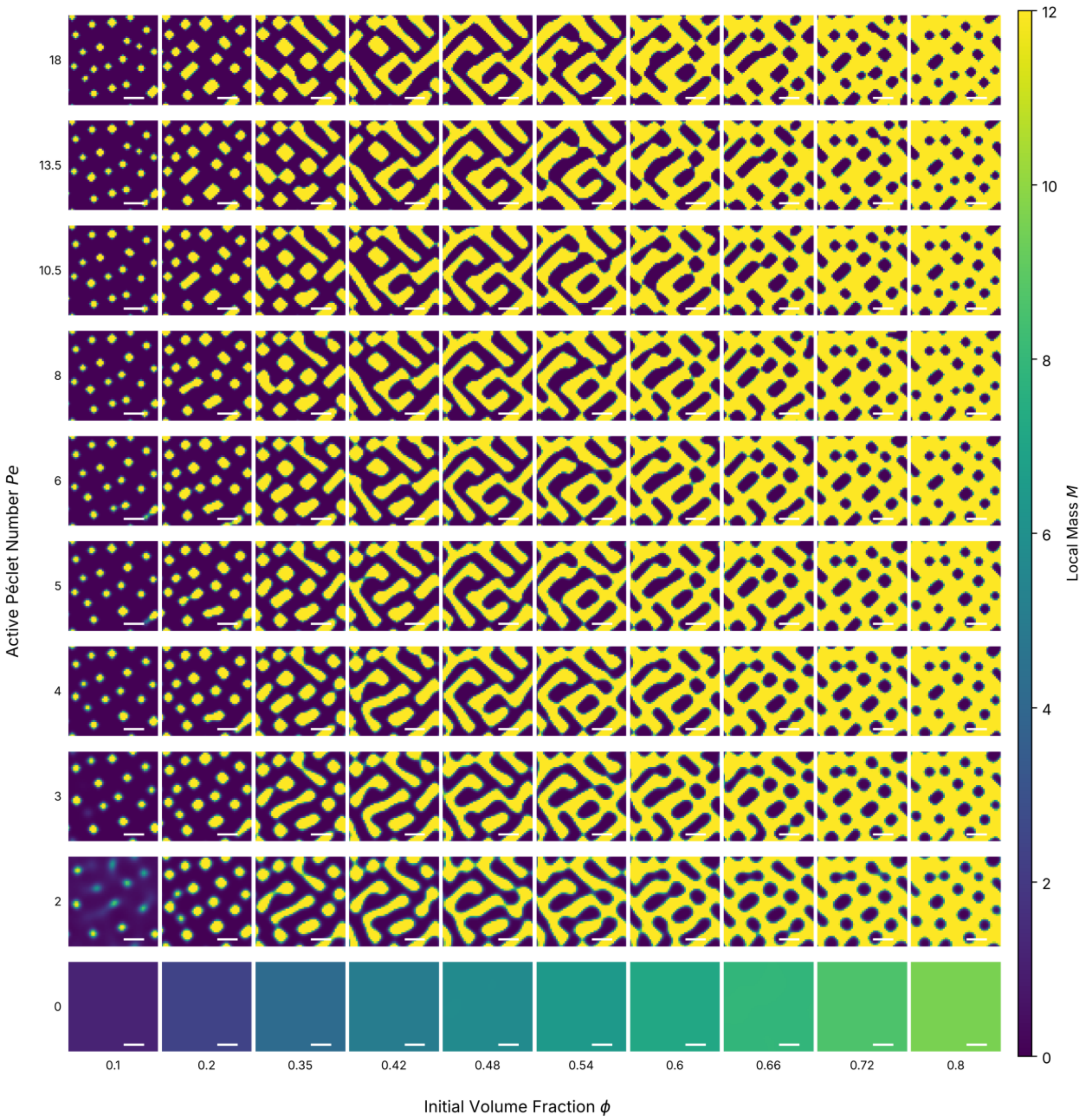

**Figure S5. Total mass concentration $M(r)$ at $\tau = 200$ across all 100 points of the $Pe, \phi$ sweep, $\gamma = 0.05$, scale bars: 10.**

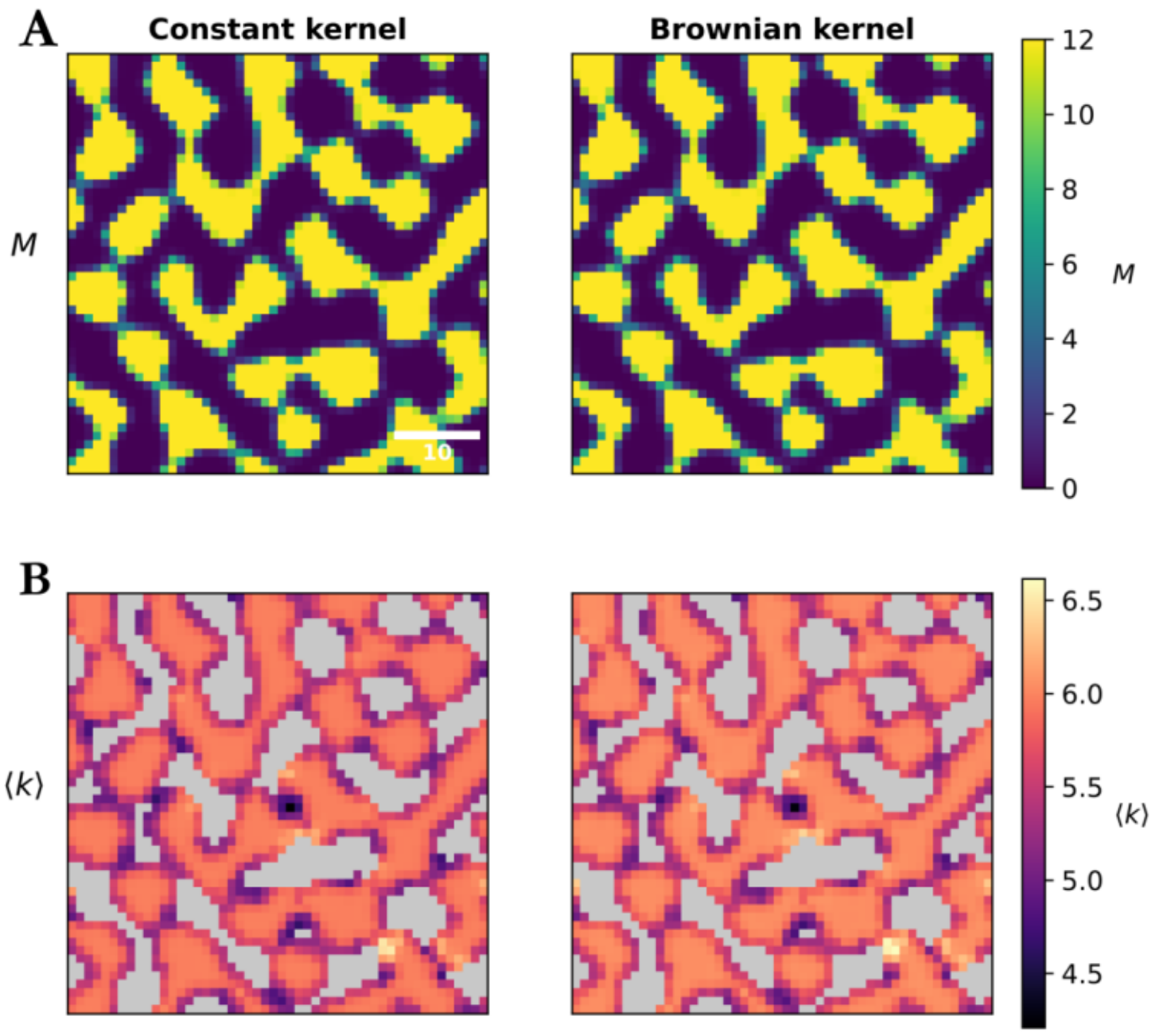

**Figure S6. Comparison of coagulation kernels ($\boldsymbol{Pe = 12, \phi = 0.45, \gamma = 0.05, K_{max} = 10}$ at $\boldsymbol{\tau = 10}$, scale bars: 10) with the constant kernel $\boldsymbol{K(i,j) = K_0}$ and the Brownian Smoluchowski kernel $\boldsymbol{K(i,j) = K_0 \frac{2+\sqrt{i/j}+\sqrt{j/i}}{4}}$, normalized so that $\boldsymbol{K(i,i) = K_0}$. (A)** Total mass concentration $M(\boldsymbol{r})$: identical. **(B)** Mean cluster size $\langle k \rangle$: same pattern, values about 1% higher with the Brownian kernel.

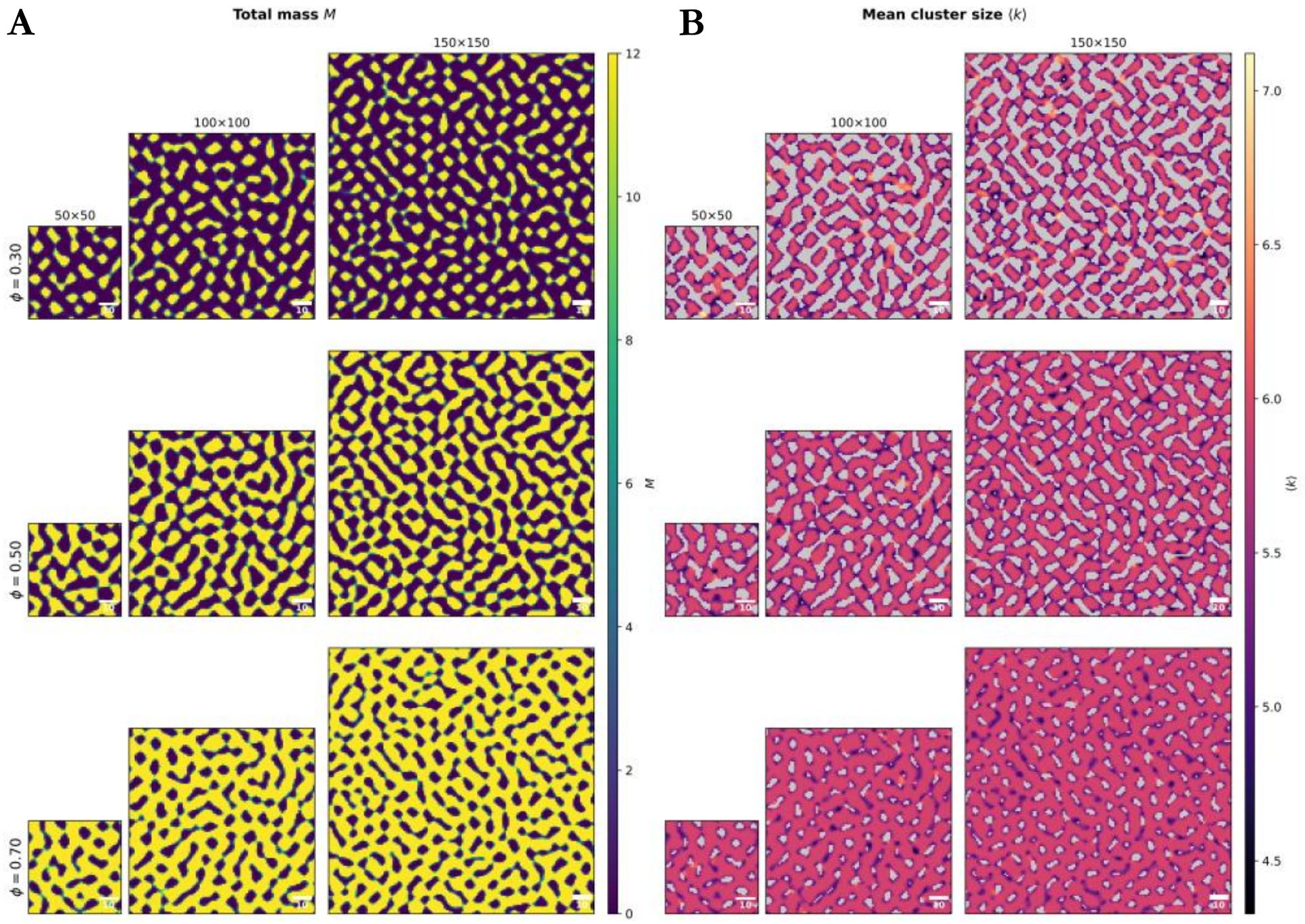

**Figure S7. (A) Total mass M and (B) mean cluster size $\boldsymbol{\langle k \rangle}$ at $\boldsymbol{\phi = 0.30, 0.50}$ *and* $\boldsymbol{0.70}$ in $\boldsymbol{50 \times 50, 100 \times 100}$ *and* $\boldsymbol{150 \times 150}$ domains, drawn at the same scale ($\boldsymbol{Pe = 12, \gamma = 0.05, K_{max} =}$**

**$10, \tau = 10$; grey cell $M < 0.1$; scale bars: 10).** Pattern wavelength and cluster composition do not depend on domain size.

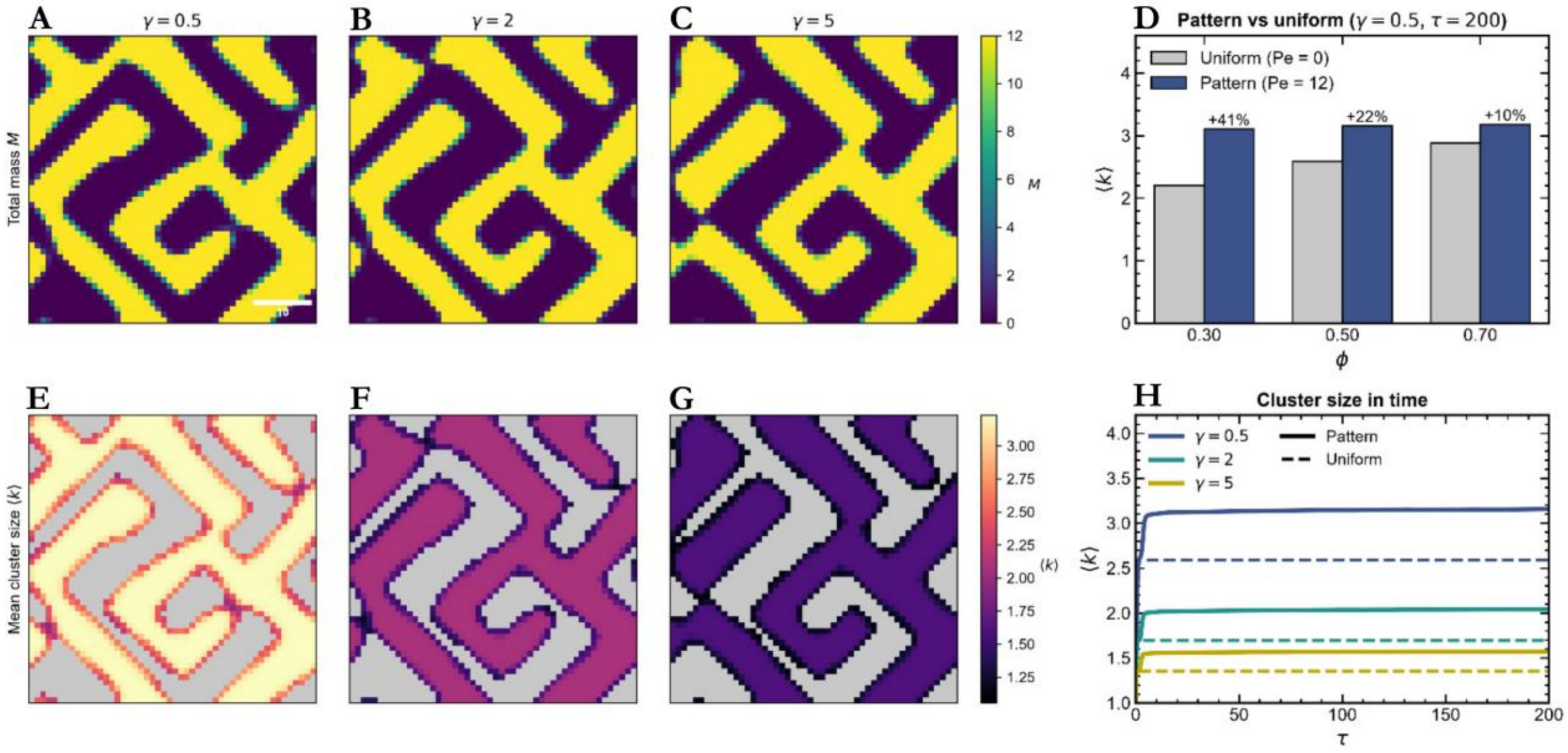

**Figure S8. Fig. 3 at long times. The runs of Fig. 3 continued to $\tau = 200$ ($Pe = 12, K_{max} = 10$, seed 42).** (**A-C**) Total mass $M(\boldsymbol{r})$ and (**E-G**) mean cluster size $\langle k \rangle$ at $\phi = 0.50$ and $\gamma = 0.5, 2$ and 5; grey cells have $M < 0.1$; scale bar: 10 lattice units. (D) Mean cluster size over the whole domain for the uniform ($Pe = 0$) and the patterned mixture ($Pe = 12$) at $\phi = 0.30, 0.50$ and $0.70$ ($\gamma = 0.5$). (**H**) Mean cluster size over the whole domain versus time, with the pattern (solid) and without it (dashed).

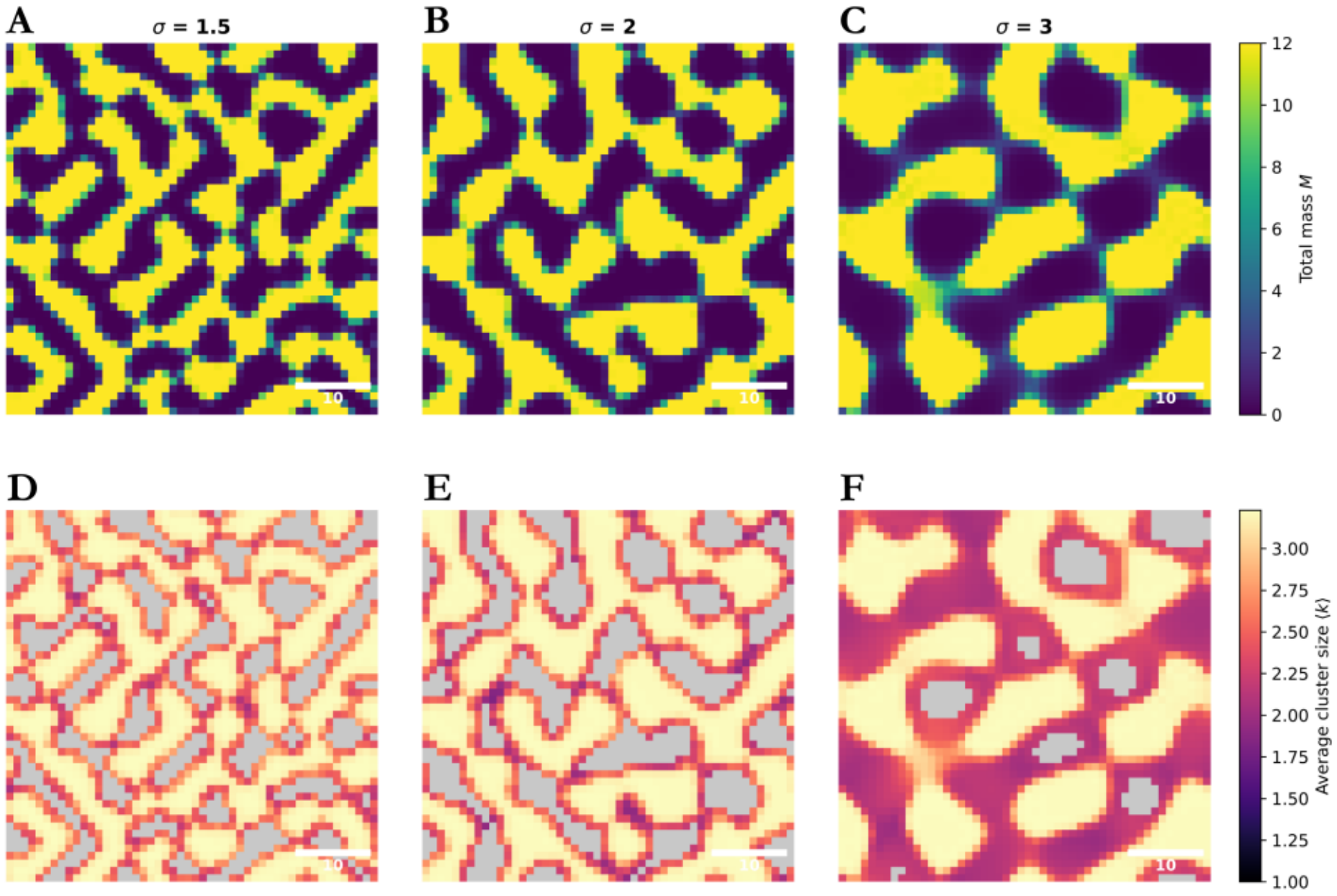

**Figure S9. Interaction range.** Total mass **M (A-C)** and mean cluster size $\langle k \rangle$ **(D-F)** for $\sigma = 1.5, 2$ and 3 at $\phi = 0.50, \gamma = 0.5, Pe = 12$ and $\tau = 10$. Grey: $M < 0.1$.

**Table S2. Robustness to noise realization and domain size.** For all the simulations we use different seeds and fixed values of $Pe = 12, \phi = 0.50, K_{max} = 10,$ and $\tau = 10$. The measured quantities are defined as follows: $\lambda$ is the pattern wavelength derived from the centroid of the $S(q)$ peak (in lattice units); the occupied area is the fraction of cells where $M > \bar{M} + \epsilon$ and $\langle k \rangle$ is the mean cluster size across the entire domain (calculated as total monomers divided by total objects).

| γ | Domain | Seeds | λ | Occupied area | $\langle k \rangle$ |
|---|---|---|---|---|---|
| 0.05 | 50 × 50 | 42, 1, 7, 123, 2024 | 9.5-9.9 | 0.475-0.481 | 5.85 |
| | 100 × 100 | 42, 7, 123 | 9.3-9.9 | 0.476-0.478 | 5.85-5.86 |
| 0.5 | 50 × 50 | 42, 1, 7, 123, 2024 | 9.6-10.0 | 0.475-0.481 | 3.10-3.11 |
| | 100 × 100 | 42, 7, 123 | 9.4-9.9 | 0.476-0.479 | 3.10-3.11 |
| 2 | 50 × 50 | 42, 1, 7, 123, 2024 | 9.7-10.1 | 0.479-0.486 | 2.01 |
| | 100 × 100 | 42, 7, 123 | 9.4-10.0 | 0.478-0.480 | 2.01 |
| 5 | 50 × 50 | 42, 1, 7, 123, 2024 | 9.7-10.2 | 0.479-0.482 | 1.55-1.56 |
| | 100 × 100 | 42, 7, 123 | 9.9-10.0 | 0.479-0.480 | 1.55 |

**Table S3. Size dependence of the drift coefficient.** Pattern and mean cluster size for three scalings of $\chi_k$ with cluster size: $\chi/\sqrt{k}$, as in the main text; $\chi$, independent of size; and $\chi\sqrt{k}$, as for attracting particles whose force grows with their mass. In all cases $D_k = D_1/\sqrt{k}, Pe = \chi/D_1 = 12, \phi = 0.50, \gamma = 0.5$, constant coagulation kernel ($K_{max} = 10, \tau = 10, 50 \times 50$ domain, seed 42). $\lambda$ is the pattern wavelength in lattice units, from the peak of the structure factor. $\Delta$ is the increase of $\langle k \rangle$ over the whole domain with respect to the uniform mixture ($Pe = 0, \langle k \rangle = 2.59$), which does not depend on $\chi_k$.

| $\boldsymbol{\chi_k}$ | $\boldsymbol{\lambda}$ | $\langle \boldsymbol{k} \rangle$, pattern | $\boldsymbol{\Delta}$(%) |
|---|---|---|---|
| $\frac{\chi}{\sqrt{k}}$ (main text) | 9.8 | 3.10 | +19.9 |
| $\chi$ (constant) | 9.9 | 3.11 | +20.0 |
| $\chi\sqrt{k}$ | 9.9 | 3.10 | +19.9 |

**Table S4. Effect of the pattern on the cluster population.** Mean cluster size $\langle k \rangle$ over the whole domain (total monomers divided by total objects) for the uniform mixture without attraction ($Pe = 0$) and the patterned mixture ($Pe = 12$), at the same composition and rates ($D_k = \frac{D_1}{\sqrt{k}}, \chi_k = \frac{\chi}{\sqrt{k}}, K_{max} = 10, \tau = 10, 50 \times 50$ domain, seed 42). The increase is the relative difference between the two.

| $\boldsymbol{\gamma}$ | $\boldsymbol{\phi}$ | ⟨k⟩, uniform (Pe = 0) | $\langle \boldsymbol{k} \rangle$, pattern (Pe = 12) | $\boldsymbol{\Delta}(\%)$ |
|---|---|---|---|---|
| 0.05 | 0.30 | 4.56 | 5.83 | +28% |
| | 0.50 | 5.15 | 5.85 | +14% |
| | 0.70 | 5.52 | 5.87 | +6% |
| | 1.00 | 5.91 | 5.91 | +0% |
| 0.5 | 0.30 | 2.20 | 3.04 | +38% |
| | 0.50 | 2.59 | 3.10 | +20% |
| | 0.70 | 2.88 | 3.15 | +9% |
| | 1.00 | 3.22 | 3.22 | +0% |
| 2 | 0.30 | 1.48 | 1.97 | +33% |
| | 0.50 | 1.69 | 2.01 | +19% |
| | 0.70 | 1.87 | 2.04 | +9% |
| | 1.00 | 2.08 | 2.08 | +0% |
| 5 | 0.30 | 1.23 | 1.53 | +24% |
| | 0.50 | 1.35 | 1.56 | +15% |
| | 0.70 | 1.46 | 1.57 | +8% |
| | 1.00 | 1.59 | 1.59 | +0% |

**Table S5. Numerical convergence.** Mean cluster size $\langle k \rangle$ over the whole domain for the uniform ($Pe = 0$) and the patterned mixture ($Pe = 12$) at $\tau = 10$.

| $\boldsymbol{\phi}$ | $\boldsymbol{\gamma}$ | $\boldsymbol{d\tau}$ | $\boldsymbol{\Delta x}$ | ⟨k⟩, uniform | $\langle \boldsymbol{k} \rangle$, pattern | $\boldsymbol{\Delta}$(%) | $\boldsymbol{\lambda}$ | Occupied | $\boldsymbol{max\, M}$ |
|---|---|---|---|---|---|---|---|---|---|
| 0.30 | 0.05 | $10^{-3}$ | 1 | 4.56 | 5.83 | +28.0 | 9.70 | 0.307 | 12.09 |
| | | $5 \times 10^{-4}$ | 1 | 4.56 | 5.83 | +28.0 | 9.70 | 0.307 | 12.09 |
| | | $10^{-3}$ | 0.5 | 4.56 | 5.91 | +29.8 | 9.69 | 0.304 | 12.14 |
| | 0.5 | $10^{-3}$ | 1 | 2.20 | 3.04 | +38.0 | 9.76 | 0.307 | 12.14 |
| | | $5 \times 10^{-4}$ | 1 | 2.20 | 3.04 | +38.0 | 9.76 | 0.307 | 12.14 |
| | | $10^{-3}$ | 0.5 | 2.20 | 3.12 | +41.6 | 9.77 | 0.302 | 12.24 |
| | 5 | $10^{-3}$ | 1 | 1.23 | 1.53 | +24.1 | 9.87 | 0.311 | 12.06 |
| | | $5 \times 10^{-4}$ | 1 | 1.23 | 1.53 | +24.1 | 9.87 | 0.311 | 12.06 |
| | | $10^{-3}$ | 0.5 | 1.23 | 1.56 | +26.2 | 9.96 | 0.303 | 12.13 |
| 0.50 | 0.05 | $10^{-3}$ | 1 | 5.15 | 5.85 | +13.7 | 9.72 | 0.475 | 12.05 |
| | | $5 \times 10^{-4}$ | 1 | 5.15 | 5.85 | +13.7 | 9.72 | 0.475 | 12.05 |
| | | $10^{-3}$ | 0.5 | 5.15 | 5.91 | +14.9 | 9.67 | 0.484 | 12.08 |
| | 0.5 | $10^{-3}$ | 1 | 2.59 | 3.10 | +19.9 | 9.82 | 0.478 | 12.11 |
| | | $5 \times 10^{-4}$ | 1 | 2.59 | 3.10 | +19.9 | 9.82 | 0.478 | 12.11 |
| | | $10^{-3}$ | 0.5 | 2.59 | 3.15 | +21.8 | 9.76 | 0.484 | 12.20 |
| | 5 | $10^{-3}$ | 1 | 1.35 | 1.56 | +14.9 | 9.96 | 0.479 | 12.05 |
| | | $5 \times 10^{-4}$ | 1 | 1.35 | 1.56 | +14.9 | 9.96 | 0.479 | 12.05 |
| | | $10^{-3}$ | 0.5 | 1.35 | 1.57 | +15.9 | 9.92 | 0.485 | 12.12 |
| 0.70 | 0.05 | $10^{-3}$ | 1 | 5.52 | 5.87 | +6.3 | 9.42 | 0.653 | 12.04 |
| | | $5 \times 10^{-4}$ | 1 | 5.52 | 5.87 | +6.3 | 9.42 | 0.653 | 12.04 |
| | | $10^{-3}$ | 0.5 | 5.52 | 5.90 | +6.9 | 9.66 | 0.669 | 12.06 |
| | 0.5 | $10^{-3}$ | 1 | 2.88 | 3.15 | +9.3 | 9.51 | 0.656 | 12.09 |
| | | $5 \times 10^{-4}$ | 1 | 2.88 | 3.15 | +9.3 | 9.51 | 0.656 | 12.09 |
| | | $10^{-3}$ | 0.5 | 2.88 | 3.18 | +10.4 | 9.75 | 0.669 | 12.16 |
| | 5 | $10^{-3}$ | 1 | 1.46 | 1.57 | +7.6 | 9.68 | 0.660 | 12.04 |
| | | $5 \times 10^{-4}$ | 1 | 1.46 | 1.57 | +7.6 | 9.68 | 0.660 | 12.04 |
| | | $10^{-3}$ | 0.5 | 1.46 | 1.58 | +8.3 | 9.89 | 0.673 | 12.11 |

**Table S6. Interaction range.** Mean cluster size $\langle k \rangle$ over the whole domain for the uniform ($Pe = 0$) and the patterned mixture ($Pe = 12$) at $\phi = 0.50, \gamma = 0.5$ and $\tau = 10$.

| $\boldsymbol{\sigma}$ | $\langle \boldsymbol{k} \rangle$, uniform | $\langle \boldsymbol{k} \rangle$, pattern | $\boldsymbol{\Delta}$(%) | $\boldsymbol{\lambda}$ | Occupied | $\boldsymbol{max\, M}$ |
|---|---|---|---|---|---|---|
| 1.5 | 2.59 | 3.10 | +19.7 | 7.89 | 0.476 | 12.14 |
| 2 | 2.59 | 3.10 | +19.9 | 9.82 | 0.478 | 12.11 |
| 3 | 2.59 | 3.03 | +17.3 | 13.37 | 0.474 | 12.05 |

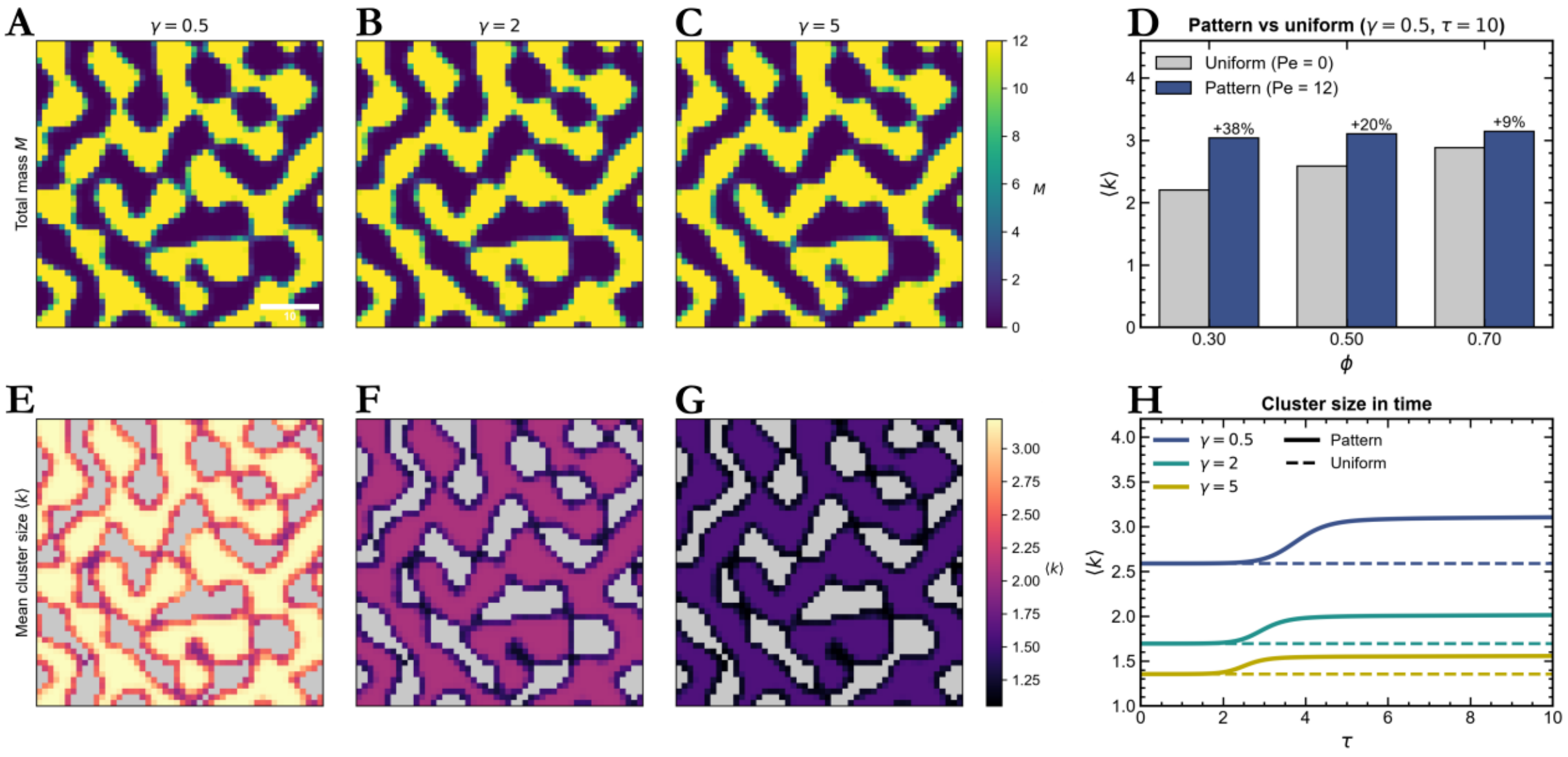

**Figure S10. Fig. 3 started from the uniform cluster distribution. The runs of Fig. 3 started from the cluster distribution of the uniform mixture at $\tau = 10$ instead of monomers, with the same total mass and noise ($Pe = 12, K_{max} = 10, \tau = 10$, seed 42). (A-C)** Total mass $M(\boldsymbol{r})$ and **(E-G)** mean cluster size $\langle k \rangle$ at $\phi = 0.50$ and $\gamma = 0.5, 2$ and 5; grey cells have $M < 0.1$; scale bar: 10 lattice units. **(D)** Mean cluster size over the whole domain for the uniform ($Pe = 0$) and the patterned mixture ($Pe = 12$) at $\phi = 0.30, 0.50$ and $0.70$ ($\gamma = 0.5$). **(H)** Mean cluster size over the whole domain versus time, with the pattern (solid) and without it (dashed). The values are the same as in Fig. 3.

## STRUCTURE FACTOR AND WAVELENGTH CALCULATION

The structure factor and characteristic pattern metrics are computed on an $L \times L$ grid with $\Delta x = 1$ using the following definitions:

i. **Mass fluctuations:**

$$\delta M(\mathbf{r}) = M(\mathbf{r}) - \langle M \rangle \qquad [S1]$$

ii. **Discrete Fourier Transform:**

$$\hat{M}(\mathbf{q}) = \sum_{\mathbf{r}} \delta M(\mathbf{r})\, e^{-i\,\mathbf{q}\cdot\mathbf{r}} \qquad [S2]$$

$$\mathbf{q} = \Delta q\,(m_x, m_y),\ m_{x,y} = -\frac{L}{2}, \dots, \frac{L}{2} - 1,\ \Delta q = \frac{2\pi}{L\,\Delta x} \qquad [S3]$$

iii. **Radial shell averaging:**

$$S(q_n) = \frac{1}{N_n} \sum_{\mathbf{q}:\ \left\lfloor \frac{|q|}{\Delta q} + \frac{1}{2} \right\rfloor = n} \left|\hat{M}(\mathbf{q})\right|^2,\quad q_n = n\,\Delta q,\ n = 0, \dots, \frac{L}{2} \qquad [S4]$$

where $N_n$ is the number of wave vectors in shell $n$ (with $S(0) = 0$).

**iv. Characteristic wavelength:**

The dominant peak index $n^* = \arg\max_{n\geq 1} S(q_n)$ defines the weighted average wave number $\bar{q}$ and wavelength $\lambda$:

$$\bar{q} = \frac{\sum_{n=n^*-2}^{n^*+2} q_n \; S(q_n)}{\sum_{n=n^*-2}^{n^*+2} S\,(q_n)}, \qquad \lambda = \frac{2\pi}{\bar{q}} \qquad [S5]$$